# Optical device for the precision control of the electric field in the focus of a beam

J. Strohaber[1]

[1]Department of Physics and Astronomy, Florida A&M University, Tallahassee, FL, 32307, USA
*james.strohaber@famu.edu*

**Abstract:** We present an inline optical device consisting of four optical components that produce a laser field consisting of two sub-beams where one is radially polarized and the other is linearly polarized. In the focus, the radially polarized sub-beam produces longitudinal polarization while the linearly polarized sub-beam produces polarization perpendicular to the propagation direction. By rotating the optical components, the orientation of the resulting electric field in the focus can be continuously varied in any direction. Estimates of the angular resolution of the device are given in the paraxial approximation.

## 1. Introduction

Since the advent of the laser in 1960, lasers have increasingly played a more important role in society. They have found practical applications in laser micro-machining [1,2], laser robotic surgery [3], telecommunications [4], and laser weaponry and defense [5]. The fundamental transverse mode of radiation emitted by most laser systems is the Gaussian or $TEM_{00}$ mode; however, more exotic forms of radiation can be produced by modifications made to the transverse phase and amplitude of the laser radiation by utilizing devices such as spatial light modulators (SLMs) and spiral phase plates to convert them into higher-order Hermite and Laguerre Gaussian beams. In the academic setting, modifications of the spatial and temporal structure of radiation have been widely used to investigate the fundamental interaction of laser radiation with matter. Specifically, spatially structured beams have been used to study molecular alignment [6], for the optical manipulation of Bose-Einstein condensates (BECs) [7], for the coherent transfer of optical orbital angular momentum to Raman sidebands [8], for micromanipulation using optical spanners and tweezers [9], for imaging of cytoskeletal filaments using stimulated emission depletion [10], for multiplexing and cryptography [11], and for optical vortex coronagraph [12]. Recently, the group of Krolikowski has experimentally demonstrated that cylindrically polarized optical vortex beams have different ablation characteristics that depend on whether the beams were radially or azimuthally polarized [1,2]. This polarization dependency has direct implications for laser machining since different polarizations are more suitable for performing various types of laser ablative cuts. For these reasons, the precision control of laser beam characteristics such as the amplitude, phase, and polarization are recognized as being vital in achieving more robust commercial devices.

In this work, we present a method and device that produces a beam that allows for full three-dimension orientation of the electric field vector within the focus of a laser beam. The resulting beam is constructed by spatially modifying the polarization and transverse phase of a laser beam with waveplates and SLMs. The resulting beam can be used for the three-dimensional alignment of molecular systems or in laser machining applications and metrology. To the best of our knowledge, no method has been reported in the literature that achieves simultaneously superimposing beams with transverse and longitudinal polarizations in such a way as to be able to continuously and precisely orientate the electric field vector in the focus.

## 2. Setup of the inline optical device

Figure 1 shows an illustration of a device consisting of a series of four optical elements used to modify the spatial phase, amplitude, and polarization of an optical beam. The input beam impinging on the device from the left is diagonally polarized and is assumed to have a transverse Gaussian beam profile. Throughout the setup, the notation for various polarization states are denoted as follows: $|H\rangle$ = horizontal, $|V\rangle$ = vertical, $|D\rangle$ = diagonal, $|L\rangle$ = left, $|R\rangle$ = right, $|k\rangle$ = longitudinal, $|D\rangle$ = diagonal and $|RP\rangle$ = radial. The first optical element (1) is a parallel aligned liquid crystal SLM programmed that adds a spatial phase structure of $\exp[i(\ell_1\theta + \varphi_1)]$ to the component of the electric field vector along the vertical (*y*-direction) while the horizontal (*x*-direction) component remains spatially unmodified by the SLM. The phase factor $\exp[i(\ell_1\theta + \varphi_1)]$ introduced by the first SLM is typically found in beams such as optical vortices [13]. These two component sub-beams are shown separately in Fig. 1; however, they are spatially superimposed. The remaining sub-beams shown in the setup are likewise spatially superimposed but have been drawn separately to illustrate modifications made to each sub-beam by each optical element. The ringed-shaped sub-beams are those encoded with an azimuthal phase, and the disk-shaped sub-beams represent Gaussian profiles. The second element (2) is a half-waveplate having a rotation angle of $\beta$ and is used to rotate the polarization states of the two sub-beams. This rotation results in four sub-beams that are then passed through element (3), which is the second SLM. The four sub-beams have components consisting of either vertical or horizontal polarization. Element (3) adds an additional phase of $\exp[i(\ell_2\theta+\varphi_2)]$ to those sub-beams having vertical polarization while leaving the initial spatial phase of the sub-beams with horizontal polarization unchanged. The final optical element (4) is a quarter-waveplate used to convert the two vortex sub-beams into a sub-beam that is radially polarized and the two Gaussian sub-beams into a sub-beam that is diagonally polarized.

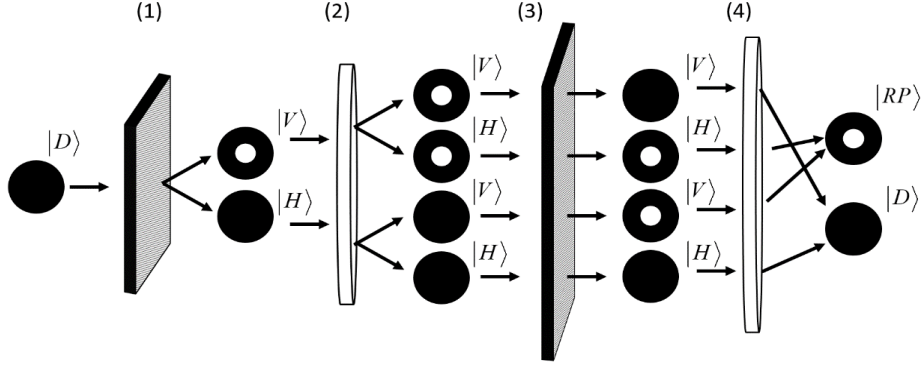

Fig. 1. Schematic representation of an inline device to orientate the electric field in the focus: The device uses a series of four optical components to convert incoming radiation into a beam in which the polarization state can be continuously varied in in 3D. Elements (1) and (3) are SLMs which spatially encode phases of $\exp[i(\ell_1\theta+\varphi_1)]$ and $\exp[i(\ell_2\theta+\varphi_2)]$ onto the beam. Element (2) is a half-waveplate having an angle of rotation of $\beta$, and element (4) is a quarter-waveplate having an angle of rotation of $\gamma$. An additional half waveplate can be positioned before the setup.

The output of the device is a beam that is simultaneously a radially polarized beam and a linearly polarized beam. This device avoids the use of an actively stabilized interferometer to superimpose radially and linearly polarized beams. In the next section, we will show that the radially polarized part, when focused, produces a longitudinal component of polarization i.e., along the propagation direction of the beam. This longitudinal component adds to the transverse linear component perpendicular to the propagation direction and results in an electric field vector that can be orientated in an arbitrary direction Fig. 2. Because the two sub-beams are never separated in the setup and because their phases and amplitudes are simultaneously controlled by waveplates and SLMs, the effect is to produce a beam where the resulting polarization is stable.

### 3. Longitudinal component of a radially polarized beam

One way to produce a beam of electromagnetic radiation with a component of the polarization along the direction of propagation requires the superposition of an even and odd Laguerre-Gaussian beam [14] with mutually perpendicular polarizations,

$$|\text{RP}\rangle = \text{LG}_{0,\ell}^{e}|\text{H}\rangle + \text{LG}_{0,\ell}^{o}|\text{V}\rangle \qquad (1)$$

The cylindrical beam in Eq. 1 is known as a radially polarized beam when the azimuthal mode number is taken to be $\ell=1$.

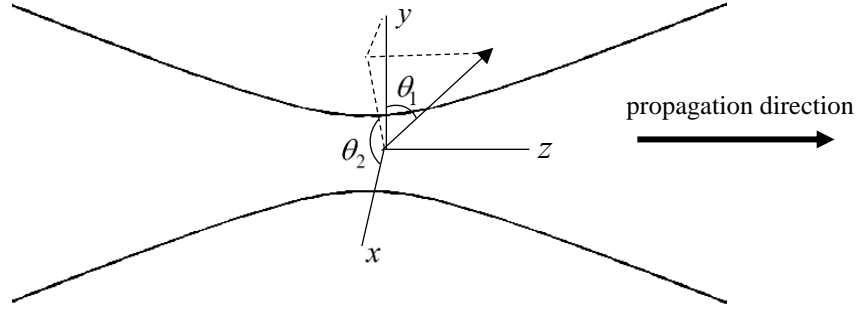

Fig. 2. Illustration of a focused laser beam. The smallest size is the waist of the beam and is at the position $z=0$. The Propagation direction of the beam is to the right and is along the $z$-axis. The $x$-axis is out of the page and the $y$-axis is vertical. The arrow shown at the center of the focus illustrates an electric field vector which can be positioned at any angle with respect to the propagation direction $\theta_1$ and in the plane perpendicular to the propagation direction $\theta_2$.

The functions $\mathrm{LG}_{0,\ell}^{(e,o)}$ are solutions of the paraxial wave equation and in cylindrical polar coordinates and are obtained by separation of variables. These solutions are known as even and odd Laguerre Gaussian beams [15],

$$\mathrm{LG}_{0,\ell}^{(e,o)} = \frac{w_0}{w}\left(\frac{\sqrt{2}r}{w}\right)^{|\ell|} \exp\left(-\frac{r^2}{w^2} - i\frac{kr^2}{2R}\right) e^{-i(|\ell|+1)\Psi_G} e^{i(kz-\omega t)} \left(\cos(\ell\theta), \sin(\ell\theta)\right) \quad (2)$$

Equation 2 can be simplified by rewriting it as the product of two functions $\mathrm{LG}_{0,\ell}^{e} = A(r,z)\cos(\ell\theta)$ and $\mathrm{LG}_{0,\ell}^{o} = A(r,z)\sin(\ell\theta)$. In the following, we will set $\ell=0$. As can be seen, Eq. 1 does not contain a longitudinal component $|k\rangle$ to the polarization. To arrive at this result, Eq. 2, which is a solution to the scalar paraxial wave equation, must satisfy the Maxwell equations [16]. By setting the magnitude of the components of the vector potential to be equal to Eq. 2 and inserting the results into the Maxwell equations, the electric field within the paraxial approximation is found to be

$$\vec{E} = E_0\left[-i(\alpha\hat{e}_x + \beta\hat{e}_y)\psi + \frac{1}{k}\left(\alpha\frac{\partial\psi}{\partial x} + \beta\frac{\partial\psi}{\partial y}\right)\hat{e}_z\right]. \quad (3)$$

Here $\psi$ can be any paraxial beam mode such as the Hermite-Gaussian (HG), Laguerre-Gaussians (LG), or Ince-Gaussians (IG) beams; and $\alpha$ and $\beta$ are polarization parameters such that when $\alpha=1$ and $\beta=0$ ($\alpha=0$ and $\beta=1$) the beam is horizontally (vertically) polarized [4]. Using the result of Eq. 3, the polarization state given in Eq. 1 can be approximated as

$$|\mathrm{RP}\rangle \approx -iA|\mathrm{RP}\rangle_\infty + \frac{1}{k}\left(\frac{\partial A}{\partial r} + \frac{A}{r}\right)|k\rangle, \quad (4)$$

where $|RP\rangle_\infty = \cos(\theta)|H\rangle + \sin(\theta)|V\rangle$ has been used. The subscript $\infty$ signifies that this is the beam outside the focus. The amplitude $A(r,z)$ has a doughnut-shaped profile and therefore has a zero in intensity along the optical axis of the beam. Close to the optical axis of the beam and in the focus, Eq. 4 can be approximated as

$$|RP\rangle \approx A(r,z)|RP\rangle_\infty + \frac{2}{k}\frac{A(r,z)}{r}|k\rangle \qquad (5)$$

The $z$-component of the polarization has a Gaussian shape with a transverse length approximately equal in size to the waist $w_0$ of the focus, and a longitudinal length approximately equal to twice the Rayleigh range $\sim 2z_0$, similar to that found for a Gaussian beam. The first term has a doughnut-shaped intensity profile with a central hole size of roughly $w_0\sqrt{\ell/2}$. Along the optical axis $r = 0$, the polarization vector in Eq. 5 is proportional to only the longitudinal polarization vector $|k\rangle$. When this field is combined with a field having a transverse polarization such as that due to a linearly polarized Gaussian beam, the superposition results in a field vector that can be varied over the two angles shown in Fig. 2. One method to achieve this is to place an s-waveplate in one arm of a stabilized interferometer having variable attenuators in each arm to control the relative powers of each sub-beam. This method is possible with current laboratory equipment but requires stabilization and for this reason, in the next section, we present an analysis of an inline setup that does not require stabilization. As a note, Eq. 4 gives the complete electric field of a radially polarized beam throughout the beam propagation, and Eq. 3 is that for a Gaussian beam, where at $z = 0$ and $r = 0$ its longitudinal component is zero.

## 4. Detailed description of the setup using Jones matrices

We use Jones vectors and matrices to describe the sub-beam polarization states and optical elements shown in Fig. 1 and determine conditions that produce the desired output beam. In the following calculations, we neglect overall phases and constants. The polarization state of the input radiation is taken to be linearly polarized at a 45 degree angle with respect to the $x$-axis $|E_1\rangle = |D\rangle$ and is written as,

$$|E_1\rangle = \frac{1}{\sqrt{2}}(|H\rangle + |V\rangle). \qquad (6)$$

This incoming polarization state can be prepared by using a half-waveplate prior to the first optical element shown in Fig. 1. This radiation is passed through element (1) where its phase structure is modified by the first SLM to introduces an overall relative phase $\varphi_1$ and a spatially varying phase of $\ell_1\theta$. The Jones matrix for this device is given by

$$M_{SLM_1} = \begin{pmatrix} 1 & 0 \\ 0 & e^{i(\ell_1\theta + \varphi_1)} \end{pmatrix}. \qquad (7)$$

The operation of this element on the beam in Eq. 6 $|E_2\rangle = M_{SLM_1}|E_1\rangle$ produces a beam encoded with the spatial phase of a Laguerre Gaussian beam.

$$|E_2\rangle = \frac{1}{\sqrt{2}}\Big(|H\rangle + e^{i(\ell_1\theta+\varphi_1)}|V\rangle\Big). \tag{8}$$

Equation 8 shows that the beam is composed of a horizontally polarized Gaussian part, and a vertically polarized part having the phase structure of an optical vortex beam. The third optical element is a half-waveplate having an arbitrary angle $\beta$ of rotation with respect to the horizontal (x-axis). The Jones matrix for this element is

$$M_{\lambda/2}(\beta) = \begin{pmatrix} \cos(2\beta) & \sin(2\beta) \\ \sin(2\beta) & -\cos(2\beta) \end{pmatrix}. \tag{9}$$

The matrix in Eq. 9 rotates the components of the input electric field vector and converts the field vector $|E_2\rangle$ given in Eq. 8 into four sub-beams,

$$|E_3\rangle = \frac{1}{\sqrt{2}}\Big(\cos(2\beta)|H\rangle + \sin(2\beta)|V\rangle + \sin(2\beta)e^{i(\ell_1\theta+\varphi_1)}|H\rangle - \cos(2\beta)e^{i(\ell_1\theta+\varphi_1)}|V\rangle\Big) \tag{10}$$

The fourth optical element is another SLM which adds an additional constant and spatially varying phase of $\varphi_2$ and $\ell_2\theta$ respectively to sub-beams polarized in the verticles direction. This is described by the same Jones matrix given in Eq. 7 with $\ell_2\theta$ and $\varphi_2$ replacing $\ell_1\theta$ and $\varphi_1$. The electric field of the radiation following this element is found to be,

$$|E_4\rangle = \cos(2\beta)\Big(|H\rangle - e^{i(\varphi_1+\varphi_2)}|V\rangle\Big) + \sin(2\beta)\Big(e^{i(-\ell_2\theta+\varphi_2)}|V\rangle + e^{i(\ell_1\theta+\varphi_1)}|H\rangle\Big) \tag{11}$$

The last optical element is a quarter-waveplate whose action, up to a constant phase, on the electric field is given by,

$$M_{\lambda/4}(\gamma) = \begin{pmatrix} \cos^2(\gamma)+i\sin^2(\gamma) & (1-i)\cos(\gamma)\sin(\gamma) \\ (1-i)\cos(\gamma)\sin(\gamma) & \sin^2(\gamma)+i\cos^2(\gamma) \end{pmatrix} \tag{12}$$

Here $\gamma$ is the angle of rotation with respect to the $x$-axis. Taking $\gamma = \pi/4$, $\varphi_1 = \pi/4$ and $\varphi_2 = 3\pi/4$, the output electric field vector is found to be,

$$|E_5\rangle = \cos(2\beta)\big(|H\rangle+|V\rangle\big) + i\sin(2\beta)\Big(e^{i\ell_1\theta}|R\rangle + e^{i\ell_2\theta}|L\rangle\Big) \tag{13}$$

The first term can be rewritten using the Jones vector for diagonal polarized light and in the case where $\ell_1 = -\ell_2 = \ell = 1$ the second term is that of radially polarized light. Making this substitution, Eq. 13 can more succinctly be written as,

$$|E_5\rangle_\infty = \cos(2\beta)|D\rangle_\infty + i\sin(2\beta)|RP\rangle_\infty \tag{14}$$

The subscript $\infty$ is used to denote that the field given in Eq. 14 is the field of the beam outside the focus. From Eqs. 4 and 5, it was shown that radially polarized radiation outside the focus produces longitudinal components of the electric field within the focus. Going from outside the focus to inside the focus, the Gaussian part $|D\rangle_\infty$ in Eq. 14 acquires a phase of $\exp(i\pi/2)$ while the radially polarized part $|RP\rangle_\infty$ acquires a phase of $\exp[i(|\ell|+1)\pi/2]$. At the focus $z=0$ and along the optical axis $r=0$, Eq. 4 becomes $|RP\rangle_\infty \to 2A|k\rangle/kr$ with $A/r = \sqrt{2}/w_0$; therefore, up to an overall phase, the output radiation of the device in the focus results in the following electric field vector

$$|E_5\rangle = \cos(2\beta)|D\rangle + \sin(2\beta)\frac{2}{k}\frac{\sqrt{2}}{w_0}|k\rangle. \tag{15}$$

Equation 15 is the electric field vector in the focus of the beam. It is the superposition of two sub-beams: one having linear transverse polarization and the other having longitudinal polarization. It can be seen that the amplitude of the linear polarized part is determined by $\cos(2\beta)$ and that of the longitudinal part is determined by $\sin(2\beta)$. In this way, changing the angle $\beta$ of the half-waveplate continuously rotates the polarization from perpendicular to longitudinal. To rotate the component of the polarization in the plane perpendicular to the propagation direction, a half-waveplate can be positioned before the device and both the device and this waveplate can be rotated together.

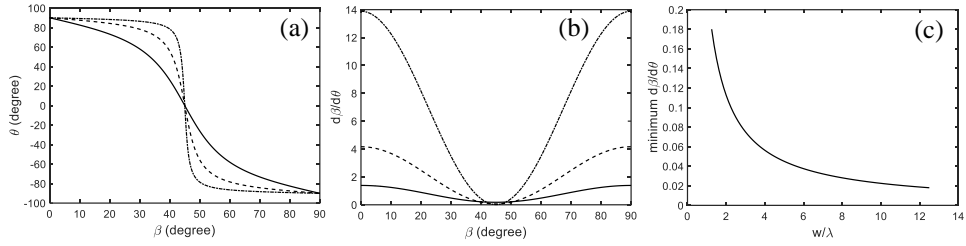

Fig. 3. (a) Tilt angle of the electric field vector within the focus as a function of waveplate angle, (b) change in waveplate angle to change in field tilt angle as a function of waveplate angle, and (c) minimum required angular step size as a function of waist to wavelength. In panels (a) and (b), the solid lines were calculated using $w_0/\lambda = 1.25$, the dashed lines with $w_0/\lambda = 3.75$ and the dash-dot lines with $w_0/\lambda = 12.5$. The minimum of the curve in panel (b) occurs for a waveplate angle of 45 degree and this minimum is shown in panel (c) for ratios of $w_0/\lambda = 1.25$ to $w_0/\lambda = 12.5$.

## 5. Estimate of the angular resolution of the device

To achieve full 3D orientation of the electric field vector in the focus requires only that the angle of rotation with the propagation direction be 90 degree since the entire device can be rotated as well. However, there are no restrictions on either of the angles shown in Fig. 2. The angle $\theta_1$ can be found from Eqs. 2, 5 and 15 and is given by,

$$\theta_1 = \arctan\left(\frac{w_0}{\lambda}\frac{\pi}{\sqrt{2}}\cot(2\beta)\right) \tag{16}$$

The resolution of this device can be estimated from the inverse of the derivative of Eq. 16,

$$\frac{d\beta}{d\theta_1} = \frac{\sqrt{2}}{2\pi} \frac{\lambda}{w_0} \left[ \left(\frac{\pi w_0}{\sqrt{2}\lambda}\right)^2 \cos^2(2x) + \sin^2(2x) \right] \quad (17)$$

From Eq. 17, the minimum estimated angular step size of the waveplate occurs when the resulting polarization vector in the focus is facing along the propagation direction which occurs at 45 degree, and the maximum angular resoltuion step size occurs at 0 degree. The minimum and maximum resolutions are found from Eq. 17 and are $d\beta/d\theta_1|_{min} = \sqrt{2}\lambda/(2\pi w_0)$ and $d\beta/d\theta_1|_{max} = \pi w_0 / 2\sqrt{2}\lambda$ respectively. Figure 3(a,b) are plots of Eqs. 16 and 17, and Fig. 3(c) is a plot of the minimum required angular step for 800 nm radiation of various waist sizes. From Fig. 3(a), it can be seen that the minimum angular step size is well within the resolution of commercially available pico-rotators.

## 6. Discussion and summary

Using an inline setup of optical elements, we have shown that it is possible to produce a beam that consists of a radially polarized and linearly polarized beam outside the focus. Upon focusing this superposition of beams, the electric field vector of the radially polarized radiation was found to tilt along the propagation direction of the beam and the linearly polarized sub-beam was perpendicular to the propagation direction. By rotating a single optical element, the magnitude of each component of the sub-beams could be adjusted allowing for the resulting electric field vector to continuously tilt between longitudinal and perpendicular polarizations while a rotation of the device together with a half waveplate preceding the device will rotate the perpendicular component of the electric field in a plane perpendicular to the propagation direction. This setup realizes smooth three-dimension orientation of the electric field within the focus of a laser beam.